\documentclass[preprintnumbers,amsmath,amssymb,floatfix,10pt,prd,onecolumn,layout,
superscriptaddress,nofootinbib]{revtex4}
\usepackage{latexsym}
\usepackage{epsfig}
\usepackage{epstopdf}
\usepackage{graphicx}
\usepackage{amssymb}
\usepackage{amsmath}
\usepackage{dcolumn}
\usepackage{bm}
\usepackage{color}
\usepackage{comment}
\usepackage{float}
\usepackage{array, makecell}
\usepackage[shortlabels]{enumitem}
\usepackage{subfigure}
\begin{document}
\title{\bf Holographic Einstein Rings of  Black Holes in Scalar-Tensor-Vector Gravity }
\author{Xiao-Xiong Zeng}
\altaffiliation{xxzengphysics@163.com}\affiliation{State Key
Laboratory of Mountain Bridge and Tunnel Engineering, Chongqing
Jiaotong University, Chongqing $400074$,
China}\affiliation{Department of Mechanics, Chongqing Jiaotong
University, Chongqing $400074$, China}
\author{M. Israr Aslam}
\altaffiliation{mrisraraslam@gmail.com}\affiliation{Department of
Mathematics, COMSATS  University Islamabad, Lahore-Campus, Lahore
$54000$ Pakistan.}
\author{Rabia Saleem}
\altaffiliation{rabiasaleem@cuilahore.edu.pk}\affiliation{Department
of Mathematics, COMSATS  University Islamabad, Lahore-Campus, Lahore
$54000$ Pakistan.}\author{Xin-Yun Hu}
\altaffiliation{ hu\_xinyun@126.com}\affiliation{College of Economic
and Management, Chongqing Jiaotong University, Chongqing $400074$,
China}
\begin{abstract}
With the help of AdS/CFT correspondence, we analyze the holographic Einstein images via the response function of the complex scalar field as a probe wave on the AdS Schwarzschild scalar-tensor-vector gravity (STVG) black hole (BH). We find that the amplitude of the response function $|\langle O\rangle|$ decreases with the increasing values of the coupling parameter $\alpha$, while it increases with the decreasing values of temperature $T$. The frequency $\omega$ of the wave source also plays a significant role in wave periods, as we increase the values of $\omega$, we find a decrease in periods of waves, which means that the total response function closely depends on the wave source. Further, we investigate the optical appearance of the holographic images of the BH in bulk. We found that the holographic ring always appears with the concentric stripe surrounded when the observer is located at the north pole, and an extremely bright ring appears when the observer is at the position of the photon sphere of the BH. With the change of the observational angle, this ring will change into a luminosity-deformed ring or a bright light spot. The corresponding brightness profiles show that the luminosity of the ring decreases, and the shadow radius increases with increasing values of $\alpha$. The relation between temperature $T$ and the inverse of the horizon $h_{e}$ is discussed, which shows the smaller values at the beginning of the horizon $h_{e}$, and then increases as the horizon radius increases. This effect can be used to distinguish the STVG BH solution from other BH solutions. Moreover, these significant features are also reflected in the Einstein ring and the corresponding brightness profiles. In addition, we compare the results obtained by wave optics and geometric optics, which are aligned well, implying that the holographic scheme adopted in this paper is valid.\\

\end{abstract}
\date{\today}
\maketitle

\section{Introduction}

During the last decades, significant progress has been made in
dealing with the strongly coupled systems owing to the
Anti-de-Sitter (AdS)/conformal field theory (CFT) correspondence
\cite{sv1,sv2,sv3}, which provides a concrete relation between
strongly coupled quantum field theories and dynamics of a classical
gravitational theory in the bulk of gravitational space-time
\cite{sv4}. This theory relates gravity in a ($d+1$)-dimensional AdS
space-time to a strongly coupled $d$-dimensional quantum field
theory living on its boundary. In this perspective, AdS/CFT
correspondence is successfully applied in different eras such as the
relation of the strong coupling dynamics of quantum chromodynamics
and the electro-weak theories, physics of black holes (BHs) and
quantum gravity, relativistic hydrodynamics, the study of condensed
matter physics, superfluidity and superconductivity
\cite{sv5,sv6,sv7,sv8,sv9,sv10,sv11,sv12,sv13,sv14,sv15}. Further,
quantum information uses AdS/CFT as a powerful tool and provides us
with an interesting result on multi-body systems, like holographic
entanglement entropy \cite{sv16}, mutual information \cite{sv17},
entanglement of purification \cite{sv18} and holographic complexity
\cite{sv19,sv20} etc.

According to AdS/CFT correspondence, we know that quantum field theories or quantum materials possess their gravity dual.  Nevertheless, no one knows about the reality of this material, especially which has its gravitational dual in our universe. However, the characteristics of such a material make it possible to analyze with classical or even quantum gravity in tabletop experiments. Hence, it is necessary to investigate the nature of such material and whether it has a gravitational dual. Recently, based on AdS/CFT correspondence, the authors in \cite{sv21} found sources on the boundary that generate one wave packet drawing a null geodesic inside the bulk. They analyzed the behavior of such a wave packet inside the bulk and built a technique that is helpful in identifying the holographic materials. In \cite{sv22}, authors proposed a method to create a star orbiting in an asymptotically AdS space-time and then they investigated the angular position of the star with the help of lensed response function in the framework of AdS/CFT correspondence. Therefore, the AdS/CFT correspondence strongly supports a better understanding of various physical topics.

On the other hand, we have witnessed the beginning of a new era in BH gravitational physics triggered by the leap in the quantity, quality, and variety of observational data from different probes. A BH is a particular astrophysical object of our Universe that possesses a powerful gravitational field, resulting in a significantly curved space-time. A BH is a solution of Einstein's field equations, describing the regions of space-time geometry that have undergone gravitational collapse. In this scenario, in 2019, the Event Horizon Telescope (EHT) provided the first strong evidence of the existence of BHs in nature and released the first ultra-high angular resolution image of the super-massive BH M$87$, which put a new spirit to further analyze the BH physics \cite{sv23,sv24}. Later, in $2021$, EHT measured the polarization of M$87$ when the magnetic fields and plasma properties are taken into account, which provides a piece of significant information about the launching of energetic jets from its core \cite{sv25,sv26}. As a milestone in the history of BH, these findings are scientifically important because these images reveal the structural information about BHs such as the accretion process of various matter, radiation mechanism, and some other relevant astrophysical consequences as well as provide strong evidence for general relativity (GR).

A prominent feature of this image is that a dark interior is surrounded by a bright ring-shaped accretion disk, in which the central dark region and the bright ring are so-called the shadow and the photon ring of the BH, respectively. The light trajectories emitted from the accretion material are deflected due to the strong gravitational field near the BH, which thus makes it possible to analyze the geometry of BH under different accretion flows. For a distant observer, the image of the BH shadow is projected onto the local sky of the observer due to gravitationally lensed, which is observed by the strong gravitational field near the BH. Hence, the study of shadow and gravitational lensing of BH are feasible methods to investigate the interesting features of the gravitational field \cite{sv27,sv28}. Since, due to different interesting scenarios of matter accretions around the BHs, the study of BH shadows and their optical appearance has reached a peak position among the scientific community. Gralla et al. \cite{sv29} analyzed the shadow and photon ring of Schwarzschild BH and found that the total luminosity of BH shadow depends on the location and form of accretion material around the BH. By establishing different accretion flow models, the study of BHs shadows has become a subject of great interest, for instance, see Refs. \cite{sv30,sv31,sv32,sv33,sv34,sv35,sv36}.

Although the properties of BH shadow and its relevant dynamics have been extensively studied in the literature, which gives us concrete information about BH physics, there are still many mysteries to uncover through some more realistic models. In this perspective, Hashimoto et al. \cite{sv37,sv38} proposed a method to construct the holographic images of the BH in the bulk, when the scalar wave emitted by the source at the AdS boundary, enters the bulk field and propagates in the bulk space-time. Particularly, they observed the Einstein ring on the holographic images and the size of the ring is consistent with the size of the BH photon sphere, which is calculated through geometrical optics. Liu et al. \cite{sv39} investigate the Einstein ring structure for the lensed response of the complex scalar field as a probe wave on the charged AdS BH in the framework of AdS/CFT correspondence. They observed that such an Einstein ring emerged at high frequencies and could be well captured by the photon sphere, away from the BH horizon, in the geometric optics approximation, and this distinct feature supports the existence of gravity dual. Further, in the framework of AdS/CFT correspondence, the authors in \cite{sv40,sv41,sv42} also investigated some distinct features of the Einstein ring in modified theories of gravity. The photon ring of the asymptotically AdS BH, dual to a superconductor, is observed on a two-dimensional sphere \cite{sv43}, where the influence of the charge scalar condensate on the image is investigated and concluded that the bulk images depicted the discontinuous change in the size of the photon ring.

Einstein's theory of GR is one of the most successful and well-established gravitational theories in modern physics, but there are many theoretical and observational shortcomings attached to it, such as the current behavior of our Universe \cite{sv44}, rotation curves of galaxies \cite{sv45} and some tension in cosmological data \cite{sv46}. In GR, we often encounter the presence of singularities and a lack of self-consistent theory of quantum gravity. In addition, from the observational point of view, GR does not address the galactic, extra-galactic, and cosmic dynamics without dark matter and dark energy \cite{sv47,sv48,sv49}. In order to resolve the problem of exotic matter, the research community divided the GR into two groups, either dark matter exists or alternatively modified Einstein's field equations. Therefore, several modified theories have been developed as the generalization of GR to resolve such problems \cite{sv50,sv51}. In \cite{sv52}, authors calculated the velocity of galaxies in clusters and found that the gravitational mass is greater than the luminous matter. In this scenario, we require modified theories, where the gravitational part is non-minimally coupled with the dynamical vector fields and hence, the dark matter/energy problem is resolved through understandable dynamical vector fields.

In literature, some promising modified gravitational theories have
been proposed from time to time, such as the Gauss-Bonnet theory
\cite{sv53}, $f(R)$ theory (in which $R$ is the Ricci scalar)
\cite{sv54} and its extended versions with minimal/non-minimal
interactions among various fields like $f(R,T)$ (where $T$ is the
trace of energy-momentum tensor) \cite{sv55}, scalar-tensor theories
and its generalized versions \cite{sv56} and teleparallel theory of
gravity \cite{sv57} etc. One of the well-known modified theories of
gravity in the literature is the so-called STVG theory proposed by
Moffat \cite{sv58}, in which the principle of action is used and
calculated through the metric tensor, three scalar fields, and a
massive vector field. In \cite{sv59}, Moffat gave the BH solution in
the context of STVG. This theory explains the cosmic phenomena
without introducing the dark matter \cite{sv60}, successfully fits
the rotation curves of galaxies \cite{sv61}, dynamic of galactic
clusters \cite{sv62}, and the cosmic microwave background (CMB)
acoustic power spectrum \cite{sv63}. In \cite{sv64}, authors
investigated the quasinormal modes of the generalized ABG
(Ay\'{o}n-Beato-Garc\'{\i}a) BHs in STVG gravity. In the framework
of the STVG model, the thermodynamics phase transition of AdS
Schwarzschild BH is investigated in \cite{sv65}. Li et al.
\cite{sv66} studied the thermal fluctuation, deflection angle, and
greybody factor of the high-dimensional Schwarzschild BH in STVG.
Cai and Miao \cite{sv67} calculated the high-dimensional static
spherically symmetric Schwarzschild BH in STVG (which is a
high-dimensional modification of STVG) and studied the quasinormal
modes of a massless scalar field and BH shadow. Recently, the shadow
of regular BH and its energy emission rate has been analyzed in STVG
theory \cite{sv68}.

In the context of AdS/CFT correspondence, in the present study, we analyze the holographic Einstein images of an AdS Schwarzschild STVG BH solutions in the bulk from a particular response function on one side of the AdS boundary, in which the response function is produced by the source, which lies far away from the response function on the other side of the AdS boundary. We employ an oscillatory Gaussian wave source $\mathcal{J}_{\mathcal{O}}$, on one side of the AdS boundary, a bulk scalar wave generated by the source, is injected into the bulk from the AdS boundary. The scalar wave propagates inside BH space-time and reaches the other side of the AdS boundary, corresponding response function will be generated such as $\langle O\rangle$, which gives the information about the bulk structure of the BH space-time \cite{sv37,sv38}. We depicted this setup in Fig. \textbf{1}.
\begin{figure}[H]\centering
\includegraphics[width=14cm,height=6.5cm]{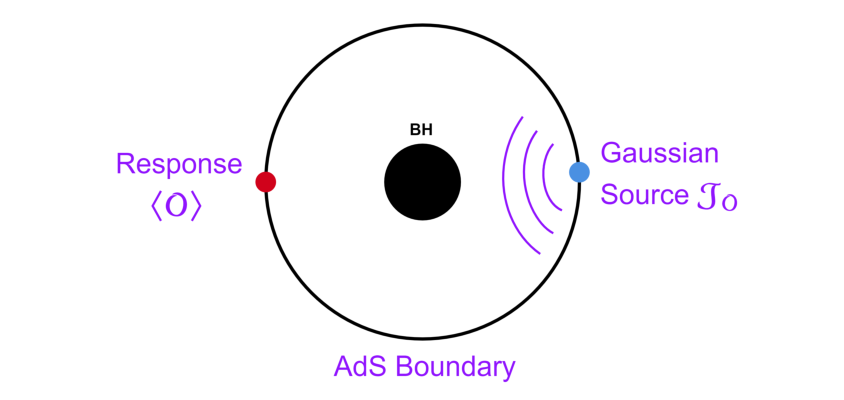} \caption{The schematic diagram of the working principle for imaging a dual BH.}
\end{figure}
Considering a special optical method, we calculate a formula, which converts the extracted response function $\langle O\rangle$, to the image of the dual BH, which can be seen on a virtual screen. The ($2+1$)-dimensional boundary CFT on a $2$-sphere $S^{2}$ is naturally dual to a BH in the AdS$_{4}$ space-time and a probe massless bulk scalar field in the space-time.

The layout of our present study is settled as follows. In section \textbf{II}, we define the summary of the AdS Schwarzschild STVG BH metric, and the holographic setup for the given response function is defined in section \textbf{III}. In section \textbf{IV}, we introduce the optical image formation in wave optics and discuss the null geodesics for the AdS Schwarzschild STVG BH on the basis of geometrical optics. We also provide a comparison of the Einstein radius of the photon sphere, which is obtained from the geometric optics with that of the holographic ring image constructed from the response function in the wave optics in the same section. The last section contains the outlook of our work.

\section{Basic Formulation of AdS Schwarzschild STVG BH}

Starting from the general action of STVG theory \cite{sv58}, which is given below
\begin{equation}\label{1}
\mathcal{I}=\mathcal{I}_{Gr}+\mathcal{I}_{\Phi}+\mathcal{I}_{S}+\mathcal{I}_{M},
\end{equation}
where
\begin{eqnarray}\label{2}
\mathcal{I}_{Gr}&=&\frac{1}{16 \pi }\int
d^4{x\sqrt{-g}}\bigg[\frac{1}{G}\bigg(R+2\Lambda\bigg)\bigg],
\\\label{3} \mathcal{I}_{\Phi}&=&-\int
d^4{x\sqrt{-g}}\bigg[\mathcal{T}\bigg(\frac{1}{4}B^{\xi\zeta}B_{\xi\zeta}+V(\Phi)\bigg)\bigg],
\\\label{4} \mathcal{I}_{S}&=&\int
d^4{x\sqrt{-g}}\bigg[\frac{1}{G^{3}}\bigg(\frac{1}{2}g^{\xi\zeta}\nabla_{\xi}G\nabla_{\zeta}G-V(G)\bigg)\bigg]+
\int
d^4{x\sqrt{-g}}\bigg[\frac{1}{G}\bigg(\frac{1}{2}g^{\xi\zeta}\nabla_{\xi}\mathcal{T}\nabla_{\zeta}\mathcal{T}-
V(\mathcal{T})\bigg)\bigg]\\\nonumber&+&\int
d^4{x\sqrt{-g}}\bigg[\frac{1}{\xi^{2}
G}\bigg(\frac{1}{2}g^{\xi\zeta}\nabla_{\xi}\xi\nabla_{\zeta}\xi-
V(\xi)\bigg)\bigg].
\end{eqnarray}
in which $\mathcal{I}_{Gr}$ is the gravity part, which is defined by the Einstein Hilbert action with additional cosmological parameter $\Lambda$, $\mathcal{I}_{\Phi}$ is the modified massive vector field of the action, $\mathcal{I}_{S}$ represents the scalar part of the action and $\mathcal{I}_{M}$ is the matter action. Further, $g$ denotes the determinant of the metric tensor, $R$ is the Ricci scalar, $V(\Phi)=\frac{1}{2}\xi^{2}\Phi^{\xi}\Phi_{\xi}$ is the self-interacting potential, $\Phi_{\xi}$ refers the massive vector field with mass parameter $\xi$ and $B_{\xi\zeta}=\partial_{\xi}\Phi_{\zeta}-\partial_{\zeta}\Phi_{\xi}$ is the anti-symmetric linear part and $\mathcal{T}$ is a dimensionless scalar field with potential $V(\mathcal{T})$. $G(x)$ and $\xi(x)$ are the scalar fields and $V(G)$ and $V(\xi)$ are the corresponding potentials, respectively. $\nabla_{\xi}$ is the covariant derivative for a metric tensor $g_{\xi\zeta}$. Since, the influence of the vector field $\Phi_{\xi}$ with mass $\xi$ becomes prominent just at kiloparsec distances from the gravitational sources, it can be neglected when solving the field equations for compact objects like BHs \cite{sv65,sv68}.

Further, at the slowly varying regime, one can consider Newton's gravity coupling parameter as $G=G_{N}(1+\alpha) $, in which $G_{N}$ is Newton's gravitational constant and $\alpha$ is a dimensionless parameter, which comes from the alternative participation of the above defined action of the scalar part $G(x)$ at the slowly varying regime. In addition, GR can be recovered when $\alpha=0$, which is usually known as the gravitational charge, so we can regard deviation of the STVG theory with respect to GR given by parameter $\alpha$. Recently, the authors in \cite{sv69} analyzed the influence of the coupling parameter $\alpha$ on the stabilization of orbits of magnetized particles moving around the Schwarzschild BH in STVG theory. The thermodynamic behavior of the Schwarzschild BH in the framework of STVG theory is discussed in \cite{sv70}, where BH heat capacity depicted the change of the sign at critical mass with the presence of Hawking radiation.

Recently, the BH solutions with a non-vanishing cosmological constant $\Lambda$ reached a peak position among the researchers due to two interesting features such as they play a significant role in the phenomenology of the AdS/CFT correspondence \cite{sv1,sv3,sv71} which associates the cosmological constant with the rank of the gauge group and the observational data shows that the universe may have a small positive value of $\Lambda$ \cite{sv72}. In fact, many authors bring our understanding to investigating the interesting features of AdS BHs in the context of different modified theories. In \cite{sv73}, the author analyzed that the cosmological parameter raises the efficiency of the Penrose process in the AdS BH for the case of $\Lambda=0$.

The line element of the AdS Schwarzschild BH in STVG theory is given as \cite{sv65}
\begin{equation}\label{5}
ds^{2}=-f(r)dt^{2}+\frac{dr^{2}}{f(r)}+r^{2}d\Omega^{2},
\end{equation}
where
\begin{equation}\label{6}
f(r)=1-\frac{2(1+\alpha)M}{r}+\frac{\alpha(1+\alpha)M^{2}}{r^{2}}+\frac{r^{2}}{l^{2}},
\end{equation}
in which $G_{N}=1$, $M$ is the mass of the BH, $l$ represents the radius of the AdS space, which is associated with the cosmological parameter as $l^{2}=-\frac{3}{\Lambda}$ for $\Lambda<0$ and $d\Omega^{2}=d\theta^{2}+\sin^{2}\theta d\phi^{2}$. Since negative $\Lambda$ in the bulk is generally related to the degrees of freedom of the dual CFT on the boundary. For simplicity, we hereafter set $l=1$. Next, we will analyze the Einstein ring structure of the above metric in the context of AdS/CFT.
\section{The Holographic Framework of AdS Schwarzschild STVG BH}
Associated with the above metric, we define the holographic mechanism of AdS Schwarzschild STVG BH in the framework of AdS/CFT correspondence. For this, we set $h=1/r$ and $f(r)=h^{-2}f(h)$. In this perspective, we define the metric (\ref{5}) in a new coordinate as following
\begin{equation}\label{7}
ds^{2}=\frac{1}{h^{2}}[-f(h)dt^{2}+\frac{dh^{2}}{f(h)}+d\Omega^{2}].
\end{equation}
For a massless scalar field, we define the Klein-Gordon equation, which is helpful in determining the dynamics, which are defined as \cite{sv38}
\begin{equation}\label{8}
\square\Psi=0.
\end{equation}
In order to solve the above equation, we define the ongoing Eddington coordinates as \cite{sv39}
\begin{eqnarray}\label{9}
v=t+h_{\star}=t-\int\frac{1}{f(h)}dh,
\end{eqnarray}
and then the metric function can be further expressed as
\begin{equation}\label{10}
ds^{2}=\frac{1}{h^{2}}[-f(h)dv^{2}-2dhdv+d\Omega^{2}].
\end{equation}
The asymptotic behavior of the scalar field close to the AdS boundary becomes
\begin{equation}\label{11}
\Psi(v,h,\theta,\phi)=\mathcal{J}_{\mathcal{O}}(v,\theta,\phi)+h\partial_{v}\mathcal{J}_{\mathcal{O}}(v,\theta,\phi)+
\frac{1}{2}h^{2}D^{2}\mathcal{J}_{\mathcal{O}}(v,\theta,\phi)+\langle
O\rangle h^{3}+\mathcal{O}(h^{4}),
\end{equation}
in which $D^{2}$ represents the scalar Laplacian on unit $\mathcal{S}^{2}$. From the AdS/CFT point of view, $\mathcal{J}_{\mathcal{O}}$ and $\langle O\rangle$ is interpreted as the external source and the corresponding expected response function in dual CFT, respectively \cite{sv74}. We consider that an axis-symmetric and monochromatic oscillating Gaussian wave packet source, which is located at the south pole of the AdS boundary, i.e., $\theta=\pi$. In this regard, we have
\begin{eqnarray}\label{12}
\mathcal{J}_{\mathcal{O}}(v,\theta)=e^{-i\omega
v}\exp\big[\frac{-(\pi-\theta)^{2}}{2\rho^{2}}\big]\frac{1}{2\pi\rho^{2}}=
e^{-i\omega v}\sum_{l=0}^{\infty} Y_{l0}(\theta)N_{lo},
\end{eqnarray}
in which $\rho$, $ Y_{l0}(\theta)$, and $N_{lo}$ interpreted the width of the Gaussian wave source, the spherical harmonics function and its coefficient, respectively. Due to the smallest value of the Gaussian tail, we only consider the case of $\rho<<\pi$. Further, the value of $N_{l0}$ can be calculated as
\begin{equation}\label{13}
N_{l0}=(-1)^{l}\bigg(\frac{(l+1/2)}{2\pi}\bigg)^{\frac{1}{2}}\exp\bigg[-\frac{(l+1/2)^{2}\rho^{2}}{2}\bigg].
\end{equation}
One can further decomposed the scalar field $\Psi(v,h,\theta,\phi)$ as
\begin{equation}\label{14}
\Psi(v,h,\theta,\phi)=e^{-i\omega
v}\sum_{l=0}^{\infty}\sum_{n=-l}^{l}Y_{l0}H_{l}(h)N_{l0},
\end{equation}
and the corresponding response function can be written as
\begin{equation}\label{15}
\langle O\rangle=e^{-i\omega v}\sum_{l=0}^{\infty}Y_{l0}\langle
O\rangle_{l} N_{lo}(\theta).
\end{equation}
From Eq. (\ref{14}), we obtain $H_{l}$, which satisfies the equation of motion
\begin{equation}\label{16}
h^{2}f(h)H''_{l}+[h^{2}f'(h)-2hf(h)+2i\omega h^{2}]H'_{l}+[-2i\omega
h-l(l+1)h^{2}]H_{l}=0,
\end{equation}
and its asymptotic behavior close to the AdS boundary can be written as
\begin{equation}\label{17}
\lim\limits_{h \to 0}H_{l}=1-i\omega
h+\frac{h^{2}}{2}(-l(1+l))+\langle
O\rangle_{l}h^{3}+\mathcal{O}(h^{4}).
\end{equation}
Clearly, Eq. (\ref{16}) has two boundary conditions for function $H_{l}$, i.e., horizon boundary condition for which $f(h)$ in Eq. (\ref{16}) vanishes, and the AdS boundary condition where  $H_{l}(0)=1$. Our main task is to solve Eq. (\ref{16}) according to these boundary conditions and obtain the function $H_{l}$, with the help of the pseudo-spectral method \cite{sv37,sv38}. In this case, we can extract $\langle O\rangle_{l}$ based on Eq. (\ref{17}) . Using the extracted $\langle O\rangle_{l}$ and Eq. (\ref{15}), one can obtain the value of the total response function as we depicted the profiles of the total response function in Figs. \textbf{2} to \textbf{4}. In these profiles the physical interference pattern arises from the scalar wave propagates in the bulk field and is diffracted by the BH.
\begin{figure}[H]\centering
\epsfig{file=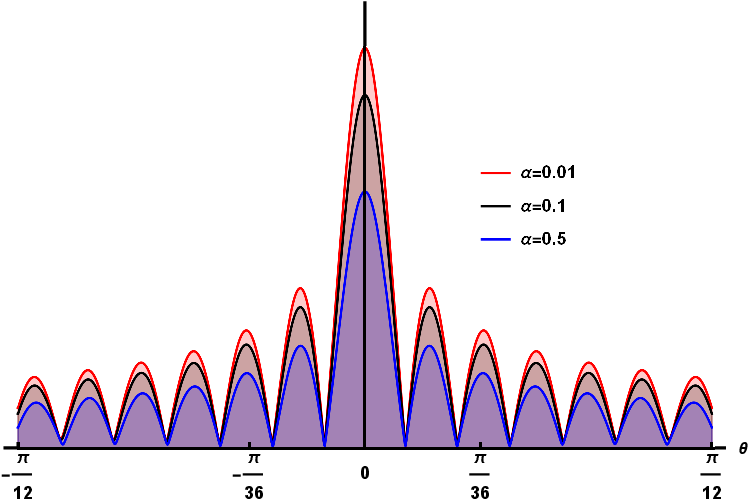, width=.5\linewidth} \caption{The amplitude of
$|\langle O\rangle|$ for different values of $\alpha$ with $h_{e}=1$
and $\omega=80$.}
\end{figure}
From Fig. \textbf{2}, clearly one can see that the amplitude of the response function shows the maximum values of the propagating wave for $\alpha=0.01$ and then decreases with the increasing values of $\alpha$. Hence, the coupling parameter of STVG theory significantly changes the behavior of the propagating waves in the bulk. We plotted Fig. \textbf{3} for different values of $\omega$, which is interpreted as the period of the oscillations of scalar waves, produced by the Gaussian source. Particularly, the period of the scalar wave is maximum when $\omega=30$, and it gradually decreases, as we increase the values of $\omega$. Further, Fig. \textbf{4} shows the amplitude of the response function for different values of Hawking temperature $T$, which is associated with the event horizon for a fixed $\alpha$. For instance, when $T=0.351$, then $h_{e}=0.8$, the amplitude of the total response function reaches the peak position. And when $T=0.568$, which corresponds to $h_{e}=0.4$, the profile of the amplitude does not significantly oscillate and shows a narrow ramp in the middle, which is hard to see, as compared to the previous one. Hence, the different values of the event horizon significantly change the behavior of the scalar waves in the bulk field, implying the temperature of the BH plays an important role in the variation of propagating waves. From the above discussion, we concluded that the optical appearance of the total response function entirely depends on the Gaussian source and the space-time geometry of the considering framework. In the next section, we further consider a special optical system, which can be converted to such a response function in the observed images, reflecting the space-time geometry more obviously.
\begin{figure}[H]\centering
\epsfig{file=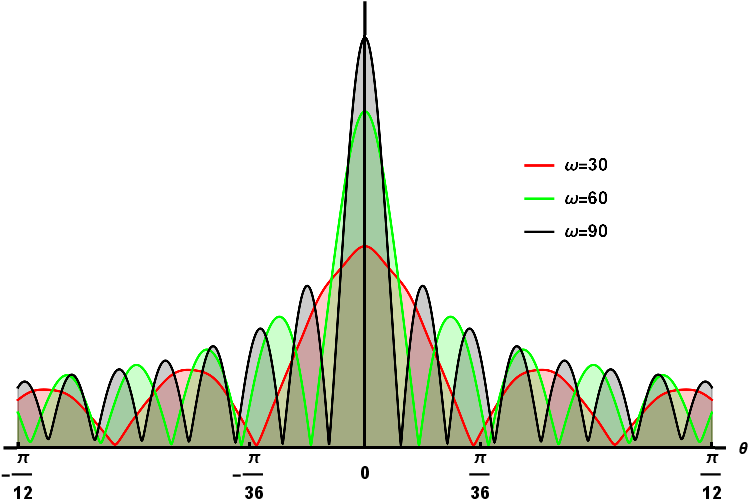, width=.5\linewidth} \caption{The amplitude period of $|\langle O\rangle|$ for different values of $\omega$ with $\alpha=0.1$ and $h_{e}=1$.}
\end{figure}
\begin{figure}[H]\centering
\epsfig{file=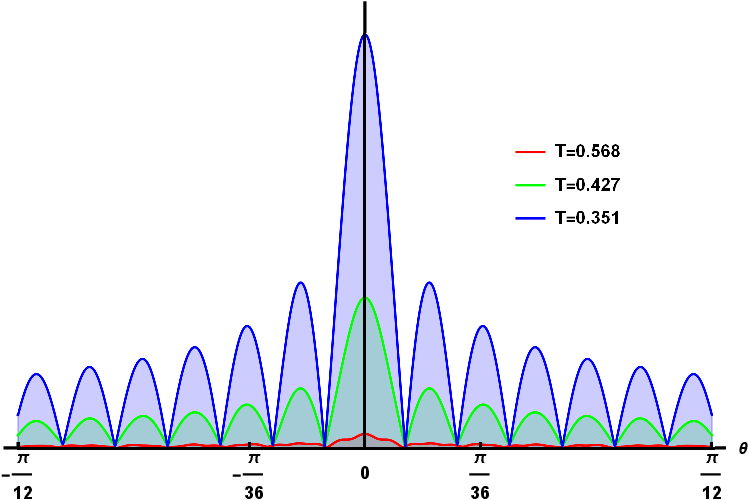, width=.5\linewidth} \caption{The amplitude of $|\langle O\rangle|$ for different values of $T$ with $\alpha=0.1$ and $\omega=80$. From top to bottom, the values of $T$ corresponds to $h_{e}=0.4,~0.6$ and $0.8$, respectively.}
\end{figure}
\section{The Optical System For Einstein Rings Formation}

After analysis of the physical interpretation of the response function, we will use it to directly interpret the image of the BH on the screen. We define a special optical system on the AdS boundary, where an observer looks up into the AdS bulk. The mechanism of our constructed optical system, which shows the image of the AdS BH from the response function, is depicted in Fig. \textbf{5}. We suppose a particular observational area at ($\theta,~\varphi)=(\theta_{obs},~0$) on the AdS boundary, where the observation region is shown in a yellow circle in the left side of Fig. \textbf{5}. Rotating the original spherical coordinate system ($\theta,~\varphi$) to a new one ($\theta',~\varphi'$) as
\begin{equation}\label{18}
\sin\theta'\cos\varphi'+i
\cos\theta'=e^{i\theta_{obs}}(\sin\theta\cos\varphi+i\cos\theta),
\end{equation}
($\theta'=0,~\varphi'=0$), which corresponds to the central point of the observational region. In addition, we define cartesian coordinates ($x,y,z$) with $(x,y)=(\theta'\cos\varphi',~\theta'\sin\varphi')$ on the AdS boundary. To construct an optical framework virtually, in the middle, we set the convex lens on the ($x,y$) plane. The focal length and radius of the convex lens are denoted by $f$ and $d$, respectively. Moreover, we adjust a spherical screen at $(x,y,z)=(x_{sr},y_{sr},z_{sr})$, satisfying $x^{2}_{sr}+y^{2}_{sr}+z^{2}_{sr}=f^{2}$ \cite{sv37,sv38}. The role of the convex lens is considered a converter between the plane and spherical waves. Imagine that an incident wave coming from the left side reaches the position of the convex lens, this wave will be converted to a transmitted wave at the focus, which will be observed on the screen, as shown in the right side of Fig. \textbf{5}.
\begin{figure}[H]\centering
\includegraphics[width=15cm,height=6.8cm]{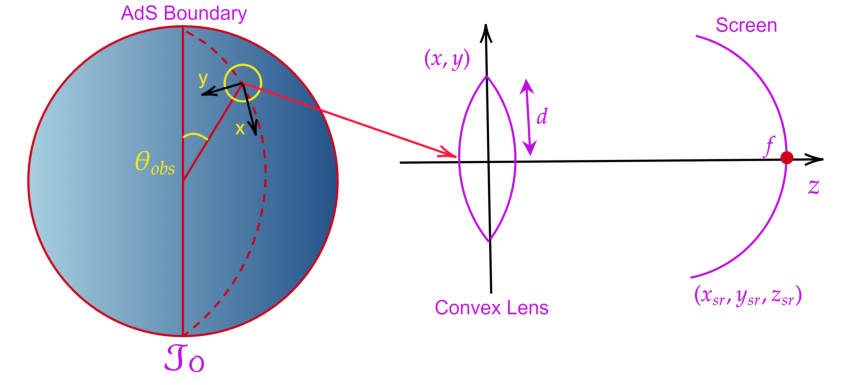} \caption{The observer's setup to see the image formed on the screen.}
\end{figure}

\begin{figure}[H]
\begin{center}
\subfigure[\tiny][~$\alpha=0.1,~\theta_{obs}=0^{o}$]{\label{a1}\includegraphics[width=3.9cm,height=4.0cm]{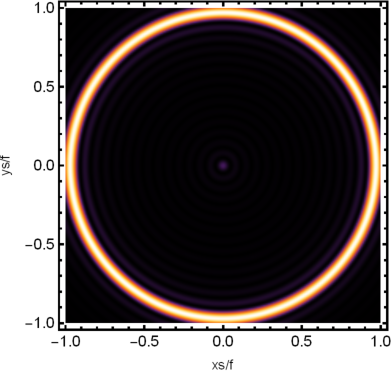}}
\subfigure[\tiny][~$\alpha=0.1,~\theta_{obs}=30^{o}$]{\label{b1}\includegraphics[width=3.9cm,height=4.0cm]{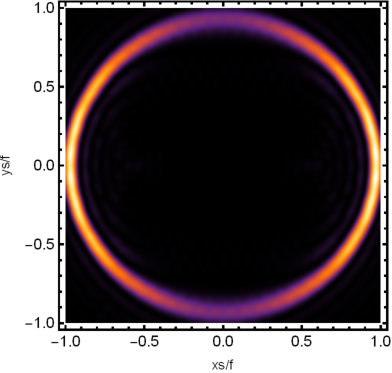}}
\subfigure[\tiny][~$\alpha=0.1,~\theta_{obs}=60^{o}$]{\label{c1}\includegraphics[width=3.9cm,height=4.0cm]{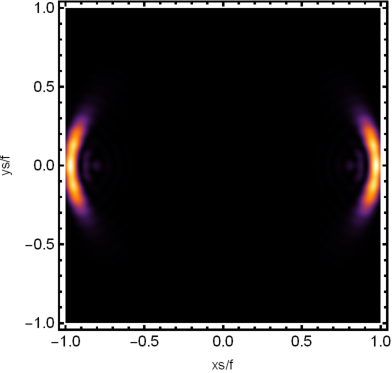}}
\subfigure[\tiny][~$\alpha=0.1,~\theta_{obs}=90^{o}$]{\label{d1}\includegraphics[width=3.9cm,height=4.0cm]{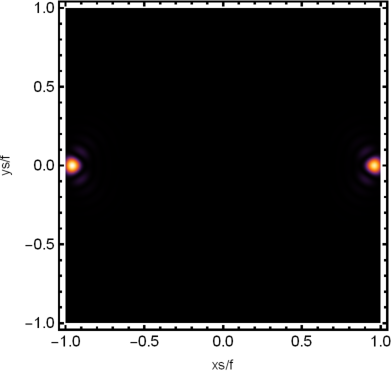}}
\subfigure[\tiny][~$\alpha=0.5,~\theta_{obs}=0^{o}$]{\label{a1}\includegraphics[width=3.9cm,height=4.0cm]{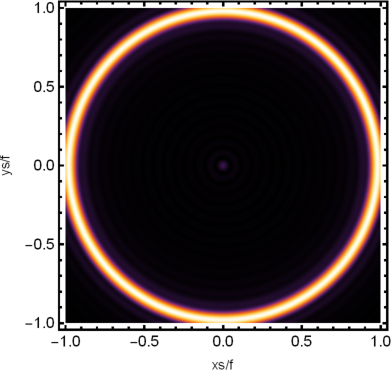}}
\subfigure[\tiny][~$\alpha=0.5,~\theta_{obs}=30^{o}$]{\label{b1}\includegraphics[width=3.9cm,height=4.0cm]{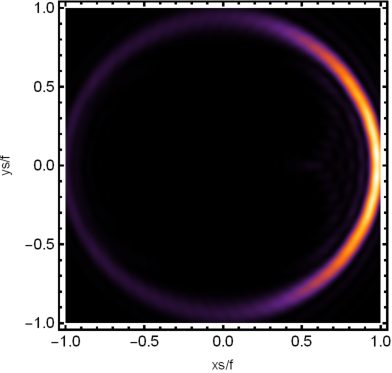}}
\subfigure[\tiny][~$\alpha=0.5,~\theta_{obs}=60^{o}$]{\label{c1}\includegraphics[width=3.9cm,height=4.0cm]{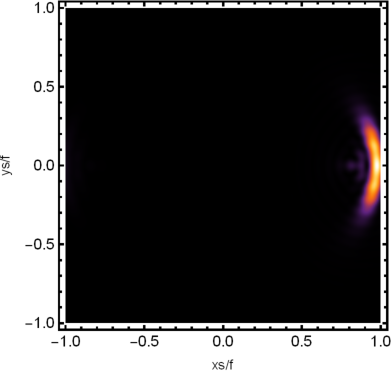}}
\subfigure[\tiny][~$\alpha=0.5,~\theta_{obs}=90^{o}$]{\label{d1}\includegraphics[width=3.9cm,height=4.0cm]{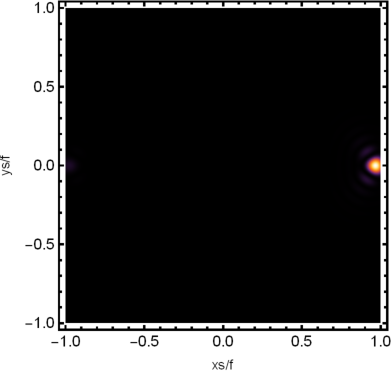}}
\subfigure[\tiny][~$\alpha=0.9,~\theta_{obs}=0^{o}$]{\label{a1}\includegraphics[width=3.9cm,height=4.0cm]{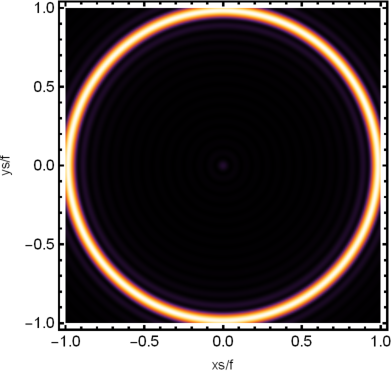}}
\subfigure[\tiny][~$\alpha=0.9,~\theta_{obs}=30^{o}$]{\label{b1}\includegraphics[width=3.9cm,height=4.0cm]{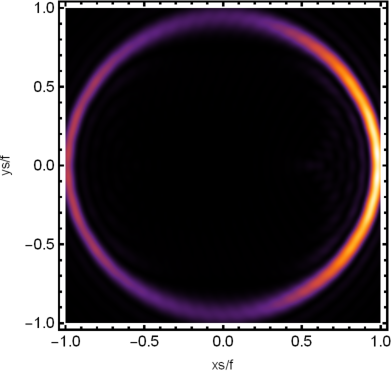}}
\subfigure[\tiny][~$\alpha=0.9,~\theta_{obs}=60^{o}$]{\label{c1}\includegraphics[width=3.9cm,height=4.0cm]{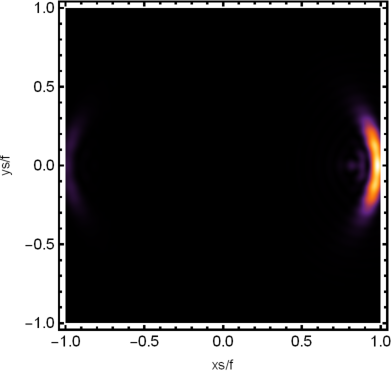}}
\subfigure[\tiny][~$\alpha=0.9,~\theta_{obs}=90^{o}$]{\label{d1}\includegraphics[width=3.9cm,height=4.0cm]{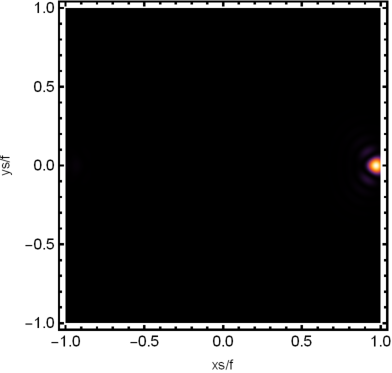}}
\subfigure[\tiny][~$\alpha=1.3,~\theta_{obs}=0^{o}$]{\label{a1}\includegraphics[width=3.9cm,height=4.0cm]{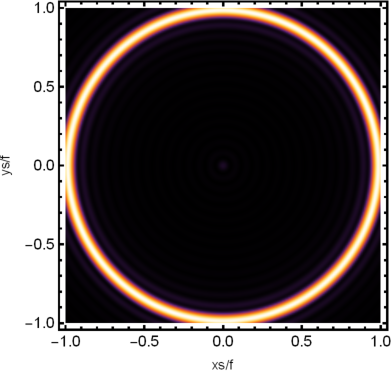}}
\subfigure[\tiny][~$\alpha=1.3,~\theta_{obs}=30^{o}$]{\label{b1}\includegraphics[width=3.9cm,height=4.0cm]{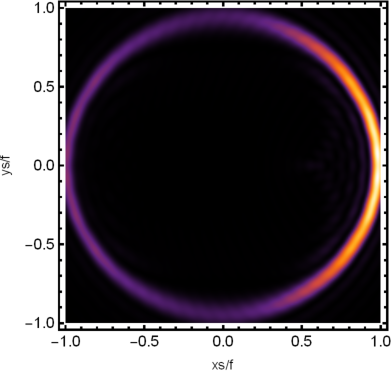}}
\subfigure[\tiny][~$\alpha=1.3,~\theta_{obs}=60^{o}$]{\label{c1}\includegraphics[width=3.9cm,height=4.0cm]{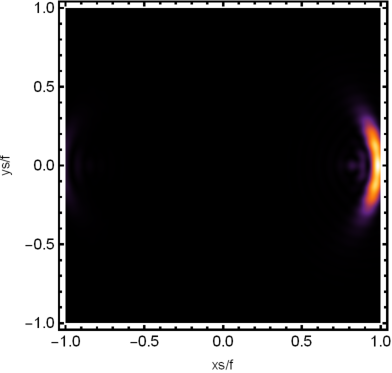}}
\subfigure[\tiny][~$\alpha=1.3,~\theta_{obs}=90^{o}$]{\label{d1}\includegraphics[width=3.9cm,height=4.0cm]{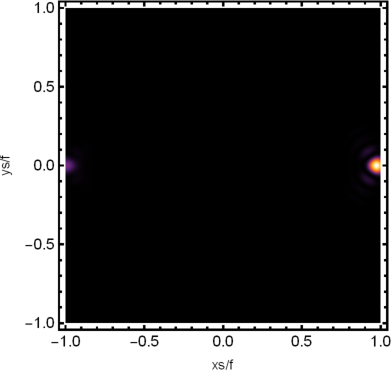}}
\caption{Observational appearance of the lensed response on the screen at various observation angles for different values of $\alpha$ with $h_{e}=1.2$ and $\omega=80$.}
 \end{center}
\end{figure}
At the point of observation, the response function is considered as a wave $\Xi(\hat{x})$, which will be transformed as the transmitted wave $\Xi_{Tr}(\hat{x})$ through the convex lens having a frequency $\omega$. Mathematically, this transmitted wave can be written as
\begin{equation}\label{19}
\Xi_{Tr}(\hat{x})=e^{-i\omega\frac{|\hat{x}|^{2}}{2f}}\Xi(\hat{x}),
\end{equation}
where $\hat{x}=(x,y,0)$ is the coordinate position on the AdS boundary. In this perspective, the optical appearance of the wave function on the screen is defined as
\begin{eqnarray}\label{20}
\Xi_{sr}(\hat{x}_{sr})=\int_{|\hat{x}|\leq
d}d^{2}x\Xi_{Tr}(\hat{x})e^{i\omega\mathbb{D}}\propto
\int_{|\hat{x}|\leq
d}d^{2}x\Xi(\hat{x})e^{-i\frac{\omega}{f}\hat{x}.\hat{x}_{sr}}=\int
d^{2}x\Xi(\hat{x})\Pi(\hat{x})e^{-i\frac{\omega}{f}\hat{x}.\hat{x}_{sr}},
\end{eqnarray}
in which $\hat{x}_{sr}$ represents the cartesian-like coordinates on the screen, $\mathbb{D}$ is the distance between $\hat{x}$ and $\hat{x}_{sr}$ and $\Pi(\hat{x})$ is the window function, which is defined as
\begin{equation}\label{21}
\Pi(\hat{x})=
    \begin{cases}
     \text{$1,$ \quad $0\leq|\hat{x}|\leq d$}, \\
     \text{$0$,\quad ~$|\hat{x}|>d$}.
    \end{cases}
\end{equation}

With the help of Eq. (\ref{20}), we can observe the image on the screen, which is obtained through the Fourier transformation of the incident wave \cite{sv75,sv76,sv77}. We depicted different profiles of the dual BH on the screen when the observer is located at different positions of the AdS boundary. For instance, we plotted the holographic Einstein images for various values of parameter $\alpha$ and observational angles for a fixed value of $h_{e}=1$ and $\omega=80$, as shown in Fig. \textbf{6}. When $\alpha=0.01$ and $\theta=0^{o}$, meaning that the observer is located at the north pole of the AdS boundary, here the observer will see a clear ring with a series of concentric striped patterns as shown in the left column of Fig. \textbf{6} (top row). These patterns are caused by a diffraction effect with imaging and are related to the lens radius $d$ and the frequency $\omega$. As we increase the values of $\alpha$ from top to bottom, we observed that the rings lie far away from the center in all cases, however the luminosity of the ring and shadow radius changes according to the values of $\alpha$, as we shown in Fig. \textbf{7}. When $\alpha=0.1$ (see the left panel of Fig. \textbf{7}), one can see that the brightness curves reached the peak position at the two endpoints on the abscissa such as $-0.97$ and $0.97$. Here, we observe the strongest brightness at these points. This effect changes when $\alpha=0.5$, where the brightness curves change the positions at the points $-0.98$ and $0.98$, it means $\alpha$ has a minor influence on the shadow radius and the brightness of the ring sharply decreases (see the second panel from the left side of Fig. \textbf{7}). Further, in the third panel of Fig. \textbf{7}, when $\alpha=0.9$, there is no change in the shadow radius but we found that the brightness of the ring gradually decreases. Further, when $\alpha$ grows like $\alpha=1.3$, we see that the position of the brightness curves is changed at the points on the abscissa $-0.99$ and $0.99$ (see the right panel of Fig. \textbf{7}).

In this case, we note that the shadow's radius is slightly increased, and the corresponding brightness of the ring gets smaller values. Hence, according to the above discussion, we found that increases of $\alpha$ lead to a change in the shadow's radius and luminosity of the ring implying that the parameter $\alpha$ is sensitive enough to analyze the luminosities and the shadow's radius of the dual BH image. Next, we fix $\theta_{obs}=30^{o}$ as the values of the parameter $\alpha$ increase, the bright ring transformed into a luminosity deformed ring and now the shape of the strict axis-symmetric ring disappears, see the second column from the left side of Fig. \textbf{6}. Further, it is noted that, when $\alpha$ has smaller values, the right side of the ring is brighter while the left side of the ring is darker. And particularly, when $\alpha=0.5$, the left side of the ring also depicted a little brightness but less than the right side of the ring. And from top to bottom, we see that the overall brightness of the ring decreases, as the values of the parameter $\alpha$ increase. Now, the concentration further goes towards the position of the angle at $\theta_{obs}=60^{o}$, see the third column from the left side of Fig. \textbf{6}.
\begin{figure}[H]
\begin{center}
\subfigure[\tiny][~$\alpha=0.1$]{\label{a1}\includegraphics[width=3.9cm,height=4.0cm]{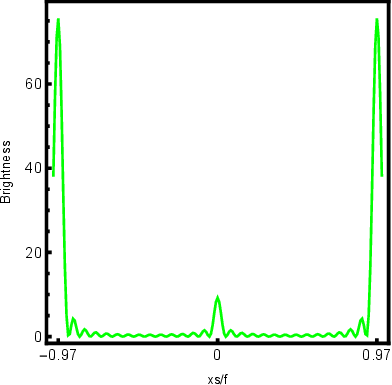}}
\subfigure[\tiny][~$\alpha=0.5$]{\label{b1}\includegraphics[width=3.9cm,height=4.0cm]{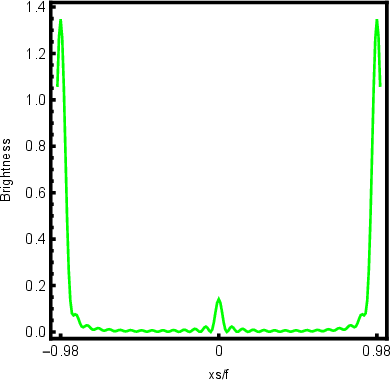}}
\subfigure[\tiny][~$\alpha=0.9$]{\label{c1}\includegraphics[width=3.9cm,height=4.0cm]{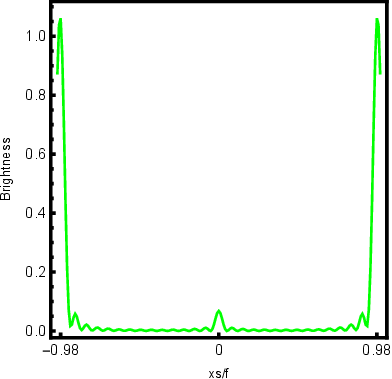}}
\subfigure[\tiny][~$\alpha=1.3$]{\label{d1}\includegraphics[width=3.9cm,height=4.0cm]{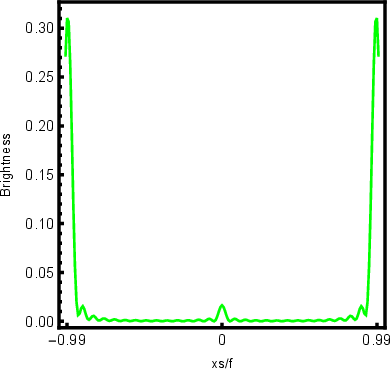}}
\caption{The brightness of the lensed response on the screen for
various $\alpha$ with $h_{e}=1.2$ and $\omega=80$.}
 \end{center}
\end{figure}
In this case, when $\alpha$ has smaller values, we see that a just bright light arcs on the right side of the screen, and when $\alpha=0.5$, a dimmer light ring also appears on the left side of the screen. From top to bottom, as the values of $\alpha$ increase, we see that there are only dimmer light arcs appearing on the left side of the screen. When we set the observation angle at $\theta_{obs}=90^{o}$, a pair of bright light spots appeared in the middle of the left and right side of the screen, as shown in the right column of Fig. \textbf{6} (top row). And when $\alpha=0.5$, there is only one bright light spot appears on the left side of the screen. As the parameter $\alpha$ increases, the bright light spot does not change the luminosity and lies far away from the center in all cases. These results are also consistent with \cite{sv41,sv42}.

Now, we are going to analyze the effect of wave source on the characteristics of the holographic Einstein image, which is observed at the position of the north pole with $\alpha=0.1$ and $h_{e}=1.2$, as shown in Fig. \textbf{8}. The corresponding profiles, which show the brightness of the lensed response on the screen for the same parameters, are shown in Fig. \textbf{9}. To see the influence of the wave source, we fix $\rho=0.02$ and $d=0.6$ for the convex lens, as the value of the frequency increases, the resulting ring becomes sharper. From this perspective, we concluded that the Einstein image that is captured through the geometric optics approximation gives a good optical appearance in the high frequency limit. This effect can also be seen in Fig. \textbf{9}, where the peak curve of brightness is getting steeper with the increasing values of the $\omega$.

\begin{figure}[H]
\begin{center}
\subfigure[\tiny][~$\omega=70$]{\label{a1}\includegraphics[width=3.9cm,height=4.0cm]{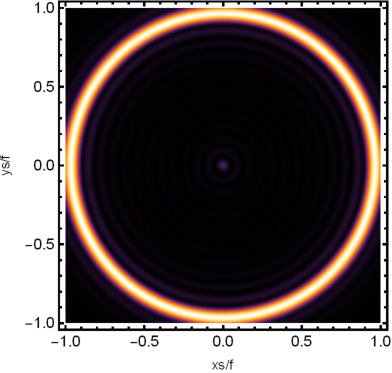}}
\subfigure[\tiny][~$\omega=50$]{\label{b1}\includegraphics[width=3.9cm,height=4.0cm]{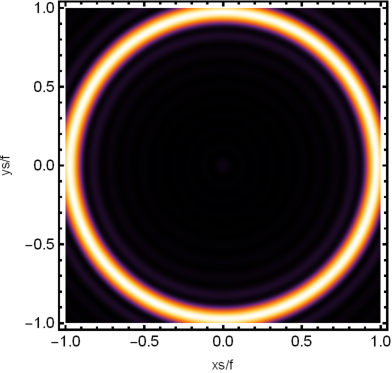}}
\subfigure[\tiny][~$\omega=30$]{\label{c1}\includegraphics[width=3.9cm,height=4.0cm]{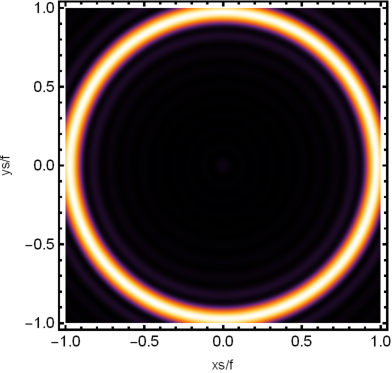}}
\subfigure[\tiny][~$\omega=10$]{\label{d1}\includegraphics[width=3.9cm,height=4.0cm]{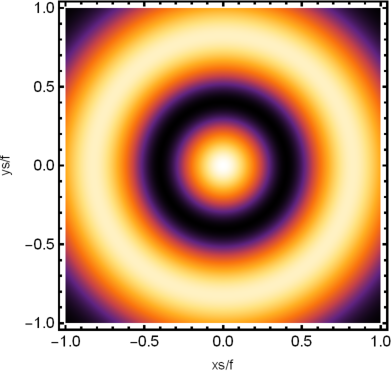}}
\caption{The profiles of the lensed response for various $\omega$ at
the observational angle $\theta=0^{o}$ with $\alpha=0.1$ and
$h_{e}=1.2$.}
 \end{center}
\end{figure}
\begin{figure}[H]
\begin{center}
\subfigure[\tiny][~$\omega=70$]{\label{a1}\includegraphics[width=3.9cm,height=4.0cm]{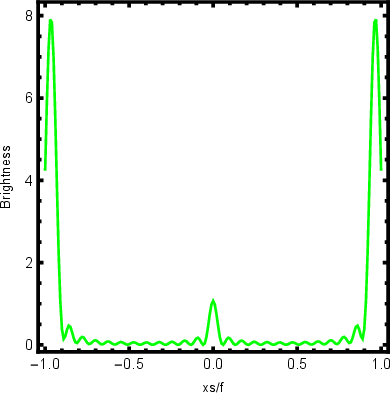}}
\subfigure[\tiny][~$\omega=50$]{\label{b1}\includegraphics[width=3.9cm,height=4.0cm]{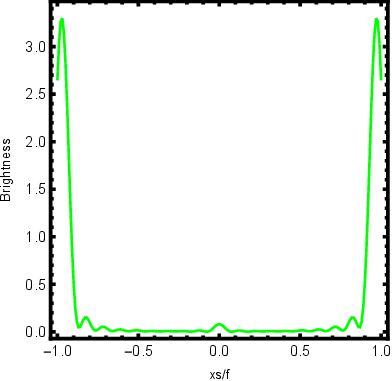}}
\subfigure[\tiny][~$\omega=30$]{\label{c1}\includegraphics[width=3.9cm,height=4.0cm]{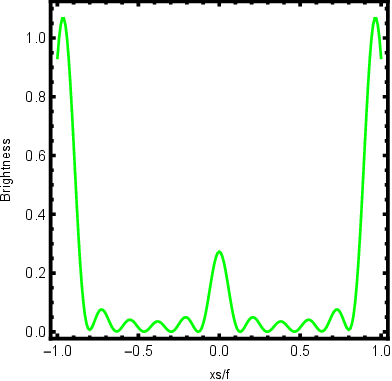}}
\subfigure[\tiny][~$\omega=10$]{\label{d1}\includegraphics[width=3.9cm,height=4.0cm]{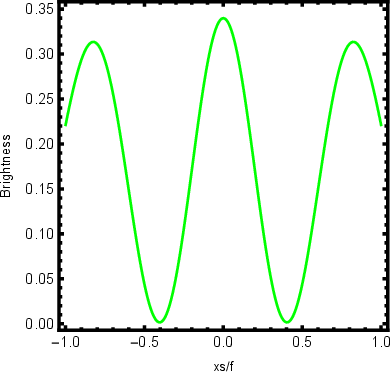}}
\caption{The brightness of the lensed response for various $\omega$ with $\alpha=0.1$ and $h_{e}=1.2$.}
 \end{center}
\end{figure}
\begin{figure}[H]\centering
\includegraphics[width=7.5cm,height=5cm]{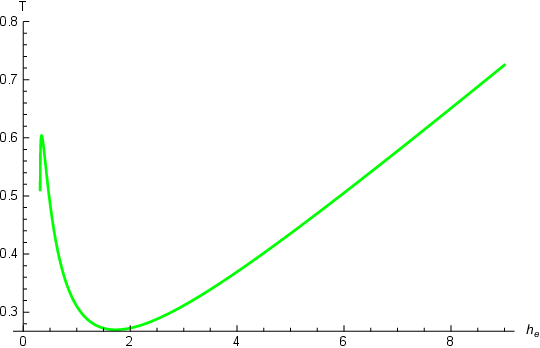} \caption{The relation between temperature $T$ and inverse of event horizon $h_e$ with $\alpha=0.1$.}
\end{figure}
We further investigate the influence of the horizon temperature $T$ on the images of the lensed response of the above holographic Einstein image, which is observed for the fixed values of the parameters $\alpha=0.1$ and the frequency $\omega=80$, as shown in Figs. \textbf{11} and \textbf{12} corresponding to small values of the horizon $h_{e}$, respectively. For a better understanding of the physical meaning of the horizon temperature, we plot the relation of temperature $T$ and the inverse of the horizon $h_{e}$ as shown in Fig. \textbf{10} with $\alpha=0.1$. It is found that the influence of temperature on shadow is different at small and large event horizons. It can be seen clearly that at the beginning the horizon temperature $T$ has the smaller value and shows the lowest value at point ($T=0.2706$, $h_{e}=1.6$). After that, it sharply increases with the increasing value of the horizon $h_{e}$. This feature may be used as a method to distinguish the STVG BH solution from other BH solutions \cite{sv39,sv40,sv42}. When the horizon $h_{e}=1.6$, we have $T=0.2706$, and similarly we have $T=0.2725,~0.3103,~0.3831$ corresponding to $h_{e}=1.5, 1, 0.7$, respectively. One can see that the horizon temperature increases with the decreasing value of the horizon $h_{e}$. In all cases, one can see that a series of axis-symmetric concentric strict rings have appeared on the screen as shown in Fig. \textbf{11}. Here, we observed that the luminosity of the rings does not change and lies far away from the center in all cases. This phenomenon can also be seen in Fig. \textbf{12}, where the brightness peak of the lensed response is far away from the center as the temperature $T$ increases for the same values of the parameters as defined in Fig. \textbf{11}. Figure \textbf{12} (see the left panel) corresponds to $T=0.2706$, illustrating that the brightness curves obtain the peak position at points $-0.96$ and $0.96$ on the abscissa and here the brightness of the ring has the maximum value.

When the temperature grows such as $T=0.2725$ (see the second panel from the left side of Fig. \textbf{12}), the shadow's radius shows a slight increase, however, the brightness of the ring decreases significantly. Similarly, as temperature increases such as $T=0.3103$ (see the third panel from the left side of Fig. \textbf{12}), we observe a further increase in the shadow's radius, but the brightness of the ring continues to decrease. When the temperature further increases and reaches $T=0.3831$ (see the right panel of Fig. \textbf{12}), the shadow's radius obtains the larger values and the brightness of the ring is still going to get smaller values. This means that increasing the horizon temperature leads to a slight increase in the shadow's radius and a decrease in the brightness of the ring. From these results, we can summarize that the smaller values of horizon $h_{e}$ are also sensitive to analyzing the optical appearance and the corresponding shadow's radius of the Einstein ring. Further, we also observe that in the middle of peak curves, there is also a relatively smaller curve of brightness, which corresponds to a very dim light spot in the center of the screen, which is hard to see in Fig. \textbf{11}.
\begin{figure}[H]
\begin{center}
\subfigure[\tiny][~$T=0.2706$]{\label{a1}\includegraphics[width=3.9cm,height=4.0cm]{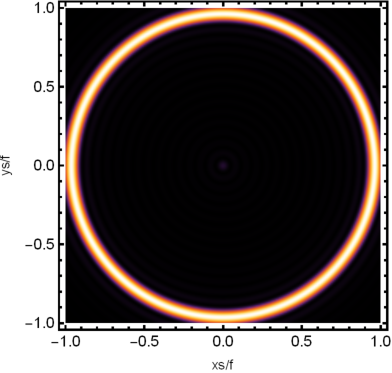}}
\subfigure[\tiny][~$T=0.2725$]{\label{b1}\includegraphics[width=3.9cm,height=4.0cm]{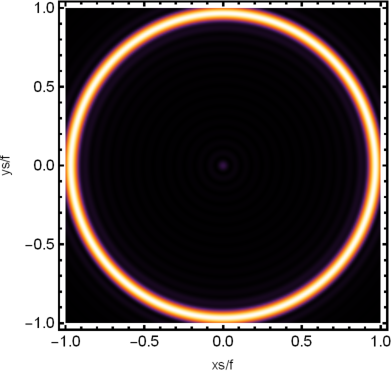}}
\subfigure[\tiny][~$T=0.3103$]{\label{c1}\includegraphics[width=3.9cm,height=4.0cm]{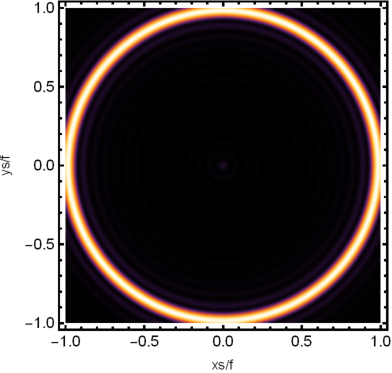}}
\subfigure[\tiny][~$T=0.3831$]{\label{d1}\includegraphics[width=3.9cm,height=4.0cm]{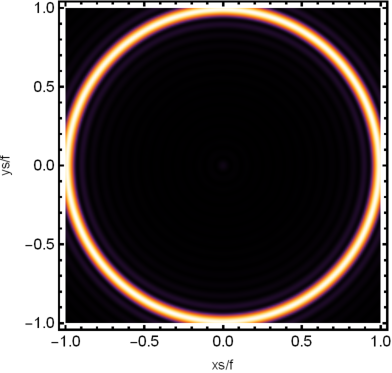}}
\caption{The profiles of the lensed response on the screen at the fixed value of the observational angle $\theta=0^{o}$ for the small $h_e$ with $\alpha=0.1$ and $\omega=80$.}
 \end{center}
\end{figure}

We further study the effect of the large values of the horizon
$h_{e}$ on the image of the dual BH as shown in Fig. \textbf{13}.
When the horizon $h_{e}=3$, we have $T=0.3113$ (see the left panel
of Fig. \textbf{13}), we observed that a series of axis-symmetric
concentric rings appeared on the screen, and one of them is a
particularly bright ring, lies far away from the center. When the
horizon grows to $h_{e}=5$ corresponding to $T=0.4354$ (see the
second panel from the left side), we see smaller axis-symmetric
concentric rings appearing in the image and the bright ring goes
closer to the center. When we further increase the value of the
horizon such as $h_{e}=7$ which corresponds to $T=0.5773$ (see the
third panel from the left side), we observe that the bright ring
comes closer to the center and ultimately the size of the ring
becomes smaller. When the horizon $h_{e}=9$ at $T=0.7252$ (see the
right panel), the brightest ring gradually moves inward. We also
plot the brightness curves of the lensed response in Fig.
\textbf{14} corresponding to the sub figures of Fig. \textbf{13}
under the same values of parameters. Particularly when $T=0.3113$,
we see that the peaks of brightness curves lie far away from the
center, and when the temperature gradually increases such as
$T=0.4354$, the peaks of the brightness curves become closer to the
center. When the temperature grows $T=0.5773$, the peak of the
brightness curves further moves towards the center. And when the
temperature rises to $T=0.7252$, the brightness curves gradually
move towards the center. Further, it is noted that in Fig.
\textbf{14}, between the peaks of the brightness curves, there is
also a relatively smaller curve of brightness, which corresponds to
the dim light spot in the center of the screen as shown in Fig.
\textbf{13}. These figures show that when the value of the
temperature is smaller, the position of the brightest ring lies far
away from the center, and at higher values, it gradually moves
inward.
\begin{figure}[H]
\begin{center}
\subfigure[\tiny][~$T=0.2706$]{\label{a1}\includegraphics[width=3.9cm,height=4.0cm]{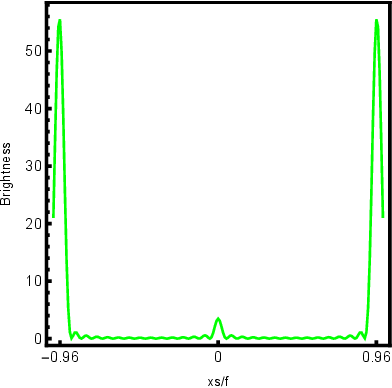}}
\subfigure[\tiny][~$T=0.2725$]{\label{b1}\includegraphics[width=3.9cm,height=4.0cm]{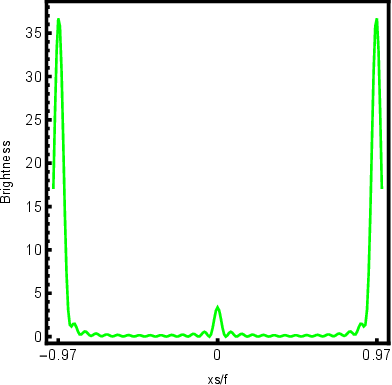}}
\subfigure[\tiny][~$T=0.3103$]{\label{c1}\includegraphics[width=3.9cm,height=4.0cm]{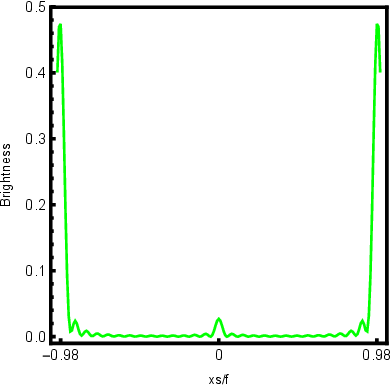}}
\subfigure[\tiny][~$T=0.3831$]{\label{d1}\includegraphics[width=3.9cm,height=4.0cm]{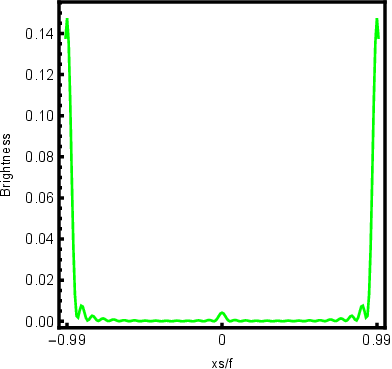}}
\caption{The brightness of the lensed response on the screen for the
small $h_e$ with $\alpha=0.1$ and $\omega=80$.}
 \end{center}
\end{figure}
\begin{figure}[H]
\begin{center}
\subfigure[\tiny][~$T=0.3113$]{\label{a1}\includegraphics[width=3.9cm,height=4.0cm]{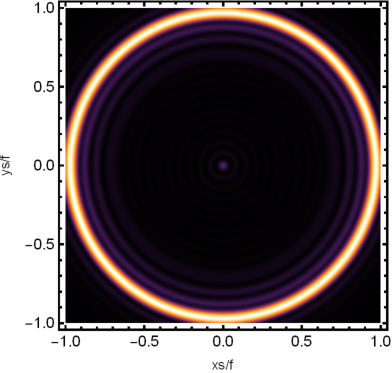}}
\subfigure[\tiny][~$T=0.4354$]{\label{b1}\includegraphics[width=3.9cm,height=4.0cm]{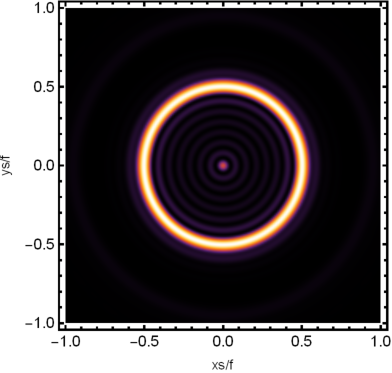}}
\subfigure[\tiny][~$T=0.5773$]{\label{c1}\includegraphics[width=3.9cm,height=4.0cm]{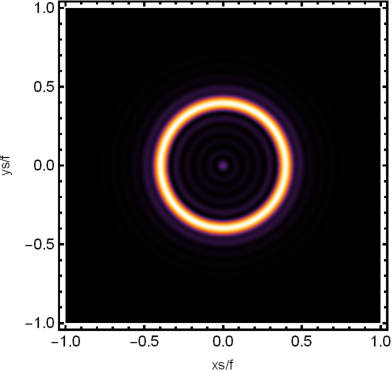}}
\subfigure[\tiny][~$T=0.7252$]{\label{d1}\includegraphics[width=3.9cm,height=4.0cm]{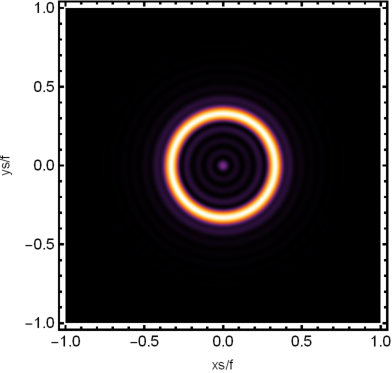}}
\caption{The profiles of the lensed response on the screen at the
fixed value of the observational angle $\theta=0^{o}$ for the large
$h_e$ with $\alpha=0.1$ and $\omega=80$.}
 \end{center}
\end{figure}

\begin{figure}[H]
\begin{center}
\subfigure[\tiny][~$T=0.3113$]{\label{a1}\includegraphics[width=3.9cm,height=4.0cm]{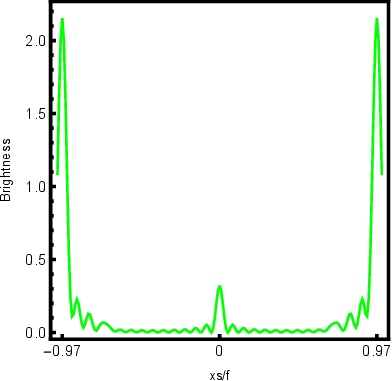}}
\subfigure[\tiny][~$T=0.4354$]{\label{b1}\includegraphics[width=3.9cm,height=4.0cm]{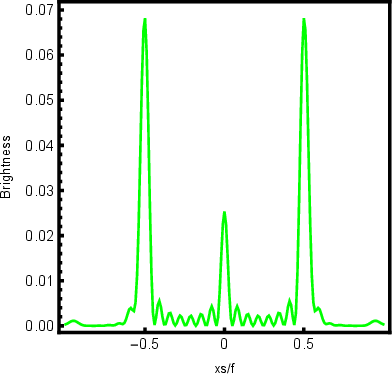}}
\subfigure[\tiny][~$T=0.5773$]{\label{c1}\includegraphics[width=3.9cm,height=4.0cm]{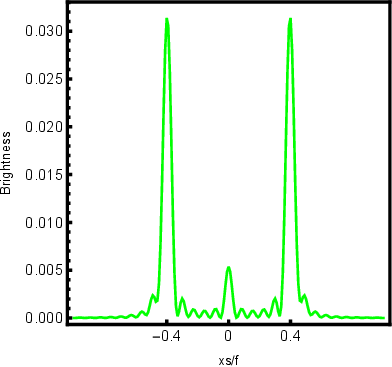}}
\subfigure[\tiny][~$T=0.7252$]{\label{d1}\includegraphics[width=3.9cm,height=4.0cm]{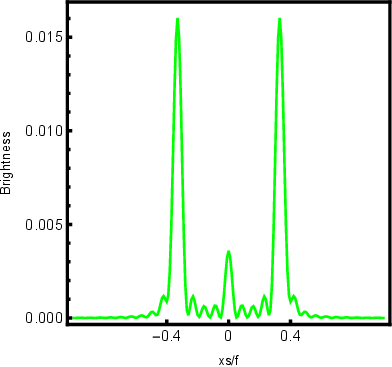}}
\caption{The brightness of the lensed response on the screen for the
large $h_e$ with $\alpha=0.1$ and $\omega=80$.}
 \end{center}
\end{figure}
In order to analyze the light deflection of AdS Schwarzschild STVG BH, we need to find how the light rays move in this space-time. Particularly, we investigate the optical geometry of the brightest ring in the image, which is located at the position of the photon sphere. In the spherically symmetric space-time, an orbit of geodesics lies in a plane passing through the center of the BH. Therefore, without loss of generality, we can always rotate an orbital plane of the null geodesic to coincide with the equatorial plane. Hence, the Lagrangian of this system is defined as \cite{sv33,sv42}
\begin{eqnarray}\label{22}
\mathcal{L}=\frac{1}{2}g_{\kappa\lambda}\dot{x}^{\kappa}\dot{x}^{\lambda}=
\frac{1}{2}\big(-f(r)\dot{t}^{2}+\frac{\dot{r}^{2}}{f(r)}+r^{2}(\dot{\theta}^{2}+\sin^{2}\theta
\dot{\phi}^{2})\big),
\end{eqnarray}
where $\dot{x}^{\kappa}$ is the four velocity of the light ray and ``.'' is the derivative with respect to the affine parameter $s$. As we consider an equatorial plane, we impose the initial conditions as $\theta=\pi/2$ and $\dot{\theta}=0$. Since the metric coefficients do not depend explicitly on time $t$ and azimuthal angle $\phi$, we have two conserved quantities, such as energy $E$ and angular momentum $L$ of the photon, defined as
\begin{eqnarray}\label{23}
E=f(r)\dot{t}, \quad L=r^{2}\dot{\phi}.
\end{eqnarray}
Based on the null geodesics, we have $g_{\kappa\lambda}\dot{x}^{\kappa}\dot{x}^{\lambda}=0$, and using Eq. (\ref{23}), we can write
\begin{equation}\label{24}
\dot{r}^{2}=E^{2}-V(r)L^{2},
\end{equation}
where $V(r)=f(r)/r^{2}$ is an effective potential. We further define the impact parameter $b=L/E$, which is the vertical distance between the geodesic line and the parallel line passing through the origin. In this space-time, there exists one null geodesic on the equatorial plane, having a circular shape. This is essentially the photon sphere, projected on the equatorial plane, yielding a ring-like circular shape, known as a photon circular orbit. Now, the effective potential $V(r)$ is used to represent the radial geodesic, which is defined as follows:
\begin{equation}\label{ef1}
\dot{r}^{2}+V(r)=\frac{1}{b^{2}}.
\end{equation}
At the photon sphere, i.e., ($r=r_{ph}$), the motions of the photons satisfy the conditions $\dot{r}=0$ and $\ddot{r}=0$, which also means that
\begin{eqnarray}\label{ef}
V(r_{ph})=\frac{1}{b^{2}_{ph}}, \quad V'(r_{ph})=0,
\end{eqnarray}
which corresponds to the maximum position of the effective potential $V(r)$. We plot the profile of the effective potential $V(r)$ in Fig. \textbf{15} for a fixed value of $\alpha$, as an example.
\begin{figure}[H]
\centering
\includegraphics[width=7.5cm,height=5cm]{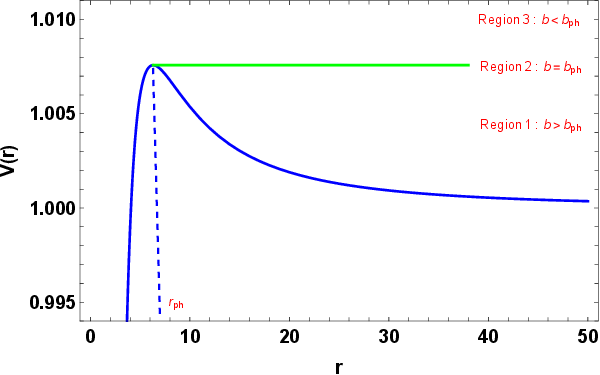}
\caption{The profile of the effective potential $V(r)$ for $\alpha=1.5$ with $M=1$. The dashed line indicates the radii of the photon sphere $r_{ph}$. The region $1$, $2$ and $3$ corresponds to $V(r)<1/b^{2}_{ph}$, $V(r)=1/b^{2}_{ph}$ (green line) and $V(r)>1/b^{2}_{ph}$, respectively.}
\end{figure}

We see that the effective potential vanishes on the event horizon. Then, it increases and reaches the peak position at the photon sphere $r_{ph}$ and subsequently moves down with respect to the radius $r$. In general, when a photon moves in the radially inward direction, it shows different behavior of motion, and we can classify it into three different regions. In region $1$, if the photon starts its motion at $r>r_{ph}$, the photon will encounter the potential barrier and be reflected in the outward direction. Further, when $b<b_{ph}$ (correspond to region $3$), the photon does not face any obstacles and hence, it will continue moving in the inward direction and finally fall into the BH singularity. Particularly, in the region $2$, namely $b=b_{ph}$, the photons start their motion at the position of the photon sphere, it will rotate around the BH in a state of constant rotation. Hence, the closer the value of the impact parameter is to $b_{ph}$, the more cycles of photons revolve around the BH.

Further, we can naturally define the ingoing angle $\theta_{in}$ of the photon with the normal vector of boundary $n^{\tau}\equiv\partial/\partial r^{\tau}$ is given as follows \cite{sv38}
\begin{eqnarray}\label{25}
\cos\theta_{in}=\frac{g_{\zeta\xi}v^{\zeta}n^{\xi}}{|v|
|n|}\bigg|_{r=\infty}=\sqrt{\frac{\dot{r}^{2}/f(r)}{\dot{r}^{2}/f(r)+L^{2}/r^{2}}}\bigg|_{r=\infty},
\end{eqnarray}
or equivalently
\begin{eqnarray}\label{26}
\sin\theta^{2}_{in}=1-\cos\theta^{2}_{in}=\frac{L^{2}V(r)}{\dot{r}^{2}+L^{2}V(r)}
\bigg|_{r=\infty}=\frac{L^{2}}{E^{2}}.
\end{eqnarray}
\begin{figure}[H]\centering
\includegraphics[width=14cm,height=7.25cm]{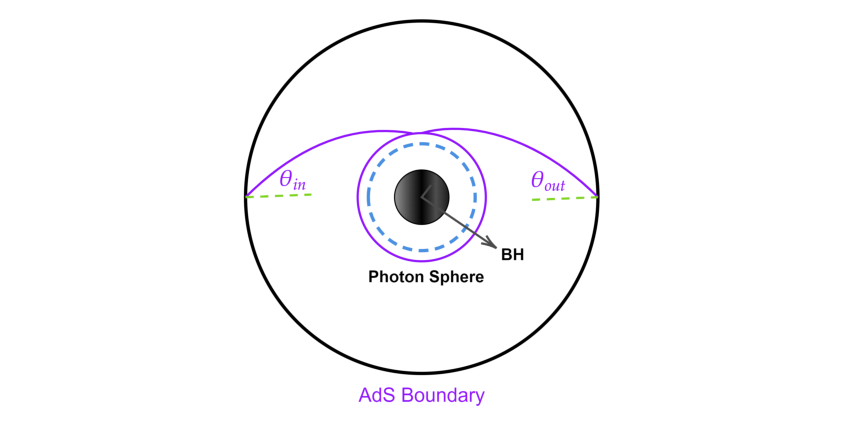} \caption{A schematic diagram of the dominant contribution to the final response function comes from the incident photon rotating once around the BH.}
\end{figure}
Hence, the ingoing angle of photon orbit from the boundary satisfies:
\begin{equation}\label{34}
\sin\theta_{in}=\frac{L}{E},
\end{equation}
which is shown in Fig. \textbf{16}. Here, the light ray approaches the photon sphere and revolves around the BH many times, since the angular velocity is non-zero. We further focused on the main contribution to the final response function, which originated from the special angular momentum as $L_{p}$, starting from the south pole, moves around the BH once and reaches the north pole, which can be determined by the following conditions as
\begin{eqnarray}\label{29}
\dot{r}=0,\quad \frac{dV}{dr}=0.
\end{eqnarray}
When an observer on the AdS boundary looks up into the AdS bulk, the resulting angle $\theta_{in}$ provides the angular distance of the image of the incident ray from the zenith. If both ends of the geodesics and the center of the BH are in alignment, the observer recognizes a picture of the ring having a radius corresponding to the incident angle $\theta_{in}$ because of axis-symmetry \cite{sv38}. In addition, as shown in Fig. \textbf{17}, one can obtain the angle of the Einstein ring, which is formed on the screen, having ring radius $\theta_{R}$ as given below
\begin{equation}\label{30}
\sin\theta_{R}=\frac{r_{R}}{f}.
\end{equation}
Further, when the angular momentum is sufficiently large such as
$\sin\theta_{in}=\sin\theta_{R}$, then we have the following
relation as \cite{sv38}
\begin{equation}\label{31}
\frac{L_{p}}{E}=\frac{r_{R}}{f}.
\end{equation}
Since, both the angle of the incident of the photon and the angle of
the photon ring describes the angle at which the observer can see
the photon ring, and hence these should be essentially equal, which
will be verified numerically, in the next. We depicted the Einstein
ring radius and the Einstein radius of the photon orbit for
different values of $\alpha$ and $\omega$ in Figs. \textbf{18} and
\textbf{19}, respectively.
\begin{figure}[H]\centering
\includegraphics[width=14cm,height=7.5cm]{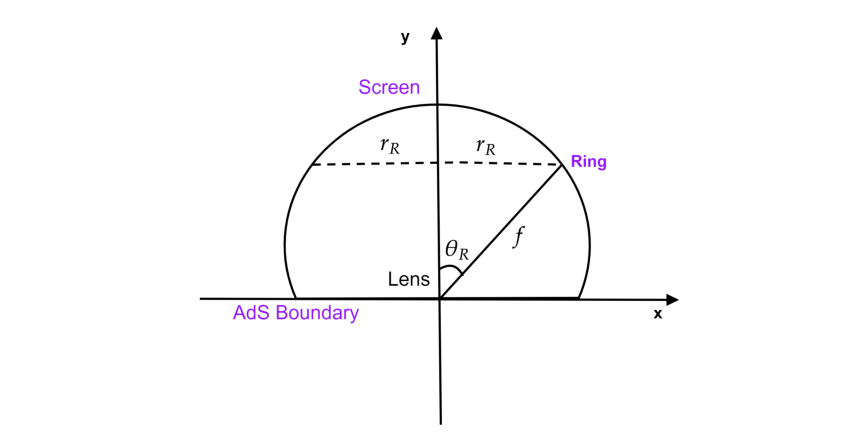} \caption{The
relation between $\theta_{R}$ and $r_{R}$.}
\end{figure}
In both Figs. the blue curve depicts the photon ring position, which
is the prediction of the geodesics, providing the angle of incidence
of the photon to the AdS boundary. Whereas the red points are the
Einstein ring radius, constructed from the response function in wave
optics, where the angular momentum can be considered as some wave
packets, realized by the superposition of the spherical harmonics.
From Fig. \textbf{18}, we observe that the Einstein ring radius
decreases with increasing temperature and the discrete red points
always lie on the blue line or in its vicinity. Further, Fig.
\textbf{19} (left panel) shows that when $\omega=80$ at the smaller
values of the temperature, the discrete red points nearly lie to the
blue line, but they gradually move away from it with increasing
values of the temperature. But as we increase $\omega$ to
$\omega=120$ (see the right panel of Fig. \textbf{19}) at the
smaller values of the temperature, the discrete red points lie
nearly below to blue line but with increasing values of temperature,
they crossed it and lie on the blue line or its vicinity. Further,
both panels of Fig. \textbf{19} illustrate that the Einstein ring
radius decreases with increasing the temperature for both values of
$\omega$. Hence, one can observe that the accuracy of high and
low-temperature fitting is different with different frequencies,
implying that the variation of $\omega$ also significantly affects
the results of both wave and geometric optics. In all cases, as
expected, the position of the photon rings given by the geometric
optics always lies on the blue line or in its vicinity with that of
wave optics.
\begin{figure}[H]
\begin{center}
\subfigure[\tiny][~$\alpha=0.1$]{\label{a1}\includegraphics[width=4.2cm,height=4.3cm]{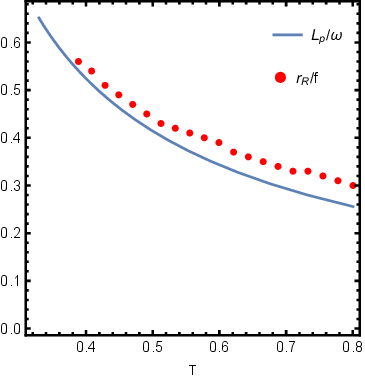}}
\subfigure[\tiny][~$\alpha=0.5$]{\label{b1}\includegraphics[width=4.2cm,height=4.3cm]{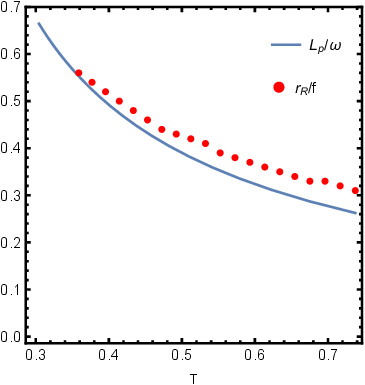}}
\subfigure[\tiny][~$\alpha=0.93$]{\label{c1}\includegraphics[width=4.2cm,height=4.3cm]{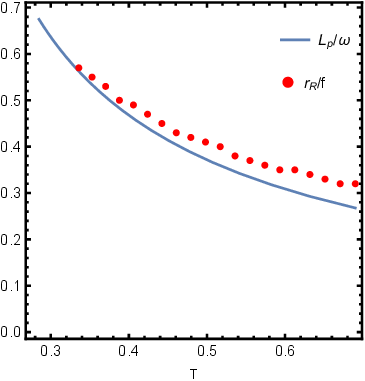}}
\subfigure[\tiny][~$\alpha=1.3$]{\label{a1}\includegraphics[width=4.2cm,height=4.3cm]{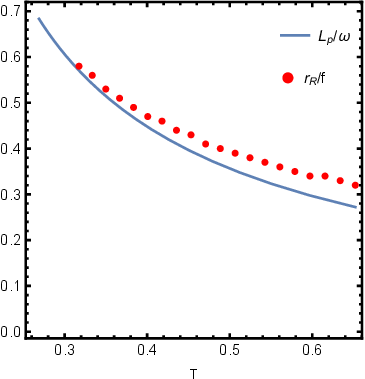}}
\caption{Comparison of Einstein ring radius (discrete red points) in the unit of $f$ with the Einstein radius of the photon orbit (blue curve) for different values of $\alpha$ with a fixed $\omega=80$.}
\end{center}
\end{figure}
\begin{figure}[H]
\begin{center}
\subfigure[\tiny][~$\omega=80$]{\label{a1}\includegraphics[width=4.2cm,height=4.3cm]{sg45.eps}}
\subfigure[\tiny][~$\omega=120$]{\label{a1}\includegraphics[width=4.2cm,height=4.3cm]{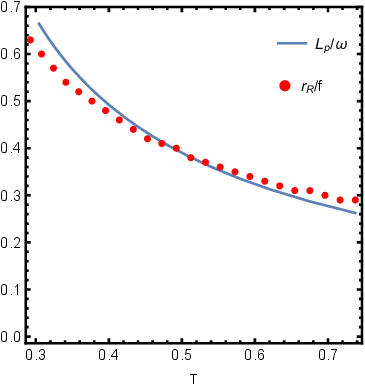}}
\caption{Comparison of Einstein ring radius (discrete red points) in the unit of $f$ with the Einstein radius of the photon orbit (blue curve) for different values of $\omega$ with a fixed $\alpha=0.5$.}
\end{center}
\end{figure}

\section{Conclusions}

Gravitational lensing by BHs is one of the most peculiar astrophysical tools for analyzing the strong field properties of gravity and provides us the information about the significant properties of distant stars which are too dim to be observed. It can help us to identify exotic objects and hence, it is fruitful to investigate the modified theories of gravity. Suppose that there is a light source behind the gravitational body. When the light source, the gravitational body, and the observer are in alignment, the observer will observe a bright ring-like shape of the light source, which is so-called the Einstein Ring. When the gravitational body is a BH, then some light rays are strongly deflected and revolve around the BH many times especially infinite times on the photon sphere. In the present study, we investigated the holographic images of a BH in the bulk from a given response function on the AdS boundary. Considering the oscillating Gaussian source $\mathcal{J}_{\mathcal{O}}$ with frequency $\omega$ produced at a point on the boundary, we derive the local response function in the context of AdS/CFT correspondence, which is presented on the other side of the AdS boundary. When the scalar wave passes through BH, it always generates a diffraction pattern in the bulk geometry of the BH space-time, which is analyzed through the local response function. Our results show that the absolute amplitude of the total response function significantly varies with the variation of coupling parameter $\alpha$, frequency $\omega$, and the positions of the horizon $h_{e}$.

As the value of $\alpha$ increases the resulting diffraction pattern decreases and this effect is vice versa when we change the values of $\omega$ and positions of the horizons. Although we analyzed the diffraction pattern of the response function, it does not directly produce concrete information on BH. With the help of an optical system, which consists of a convex lens and spherical screen, we obtained the images of the BH from the response function based on the wave optics method. We depicted the resulting Einstein ring image on the screen for different values of $\alpha$ and observational angles in Fig. \textbf{6}. When the position of the observer changes, the resulting Einstein image changes into a luminosity deformed ring, light arcs, or a bright light spot. In addition, we also show the effect of $\alpha$ on the brightness curves of the lensed response and it seems that the peak of the curves always lies far away from the center in all cases as shown in Fig. \textbf{7}. Here one can see that, when $\alpha$ has smaller values, the luminosity of the ring shows the maximum value, and it gradually decreases with the increasing values of $\alpha$. Further, in all cases, it is noted that the shadow radius increases with respect to increasing values of $\alpha$, and hence, $\alpha$ is sensitive to observing the holographic Einstein image of the STVG BH. Similarly, Figs. \textbf{8} and \textbf{9} are showing Einstein ring image and corresponding brightness profiles for different values of $\omega$. In Fig. \textbf{8}, it seems that the width of the Einstein ring becomes larger for lower values of $\omega$ and becomes sharper for large values. This effect can also be seen in Fig. \textbf{9}, where the peak curves of the brightness show sharper wave oscillations with increasing values of $\omega$. In general, the optical appearance of the Einstein ring and the brightness profiles are closely related to the BH-involved parameters, the optical system, and the wave source.

We also show the optical appearance of the horizon temperature with respect to the inverse of horizon $h_{e}$ with a fixed value of parameter $\alpha=0.1$ as shown in Fig. \textbf{10}. It is found that at the beginning of the horizon $h_{e}$, the temperature shows smaller values, and it increases with the increasing values of the horizon radius. This particular behavior of the horizon temperature may be used as a tool to distinguish the STVG BH solutions from previous studies \cite{sv39,sv40,sv42}. Further, the influence of the temperature with respect to smaller and larger values of the horizon $h_{e}$ is also depicted on the profiles of the holographic images and the brightness curves of the lensed response, as shown in Figs. \textbf{11}-\textbf{14}. It is observed that in the case of smaller values of the horizon $h_{e}$, there exists a series of concentric striped patterns in which one of them is the brightest ring, lies far away from the center, as shown in Fig. \textbf{11}. This effect can also be observed in the brightness profiles (see Fig. \textbf{12}), where the luminosity of the ring obtains smaller values as we increase the horizon temperature and the peak curves always stay close to the boundary. In addition, we found that at the smaller values of the horizon temperature, the shadow radius has minimum values, but it increases with the increasing values of the temperature. On the other hand, for large values of the horizon $h_{e}$, we found that the brightest ring gradually moves towards the center with increasing values of temperature. The proper visualization of this effect is depicted in Fig. \textbf{13}, and the corresponding brightness curves of the lensed response are shown in Fig. \textbf{14}. One can see that when $T=0.3113$, the luminosity of the ring and shadow radius show the maximum values, and the increasing values of the horizon temperature lead to a decrease in the luminosity of the ring and shadow radius (see Fig. \textbf{14}).

Based on the geometric optics, we also analyzed the Einstein ring radius corresponding to the position of the photon orbit, which is calculated from the geodesic analysis with the Einstein radius of the image constructed from the response function in the wave optics as shown in Figs. \textbf{18} and \textbf{19} for different values of parameter $\alpha$ and $\omega$, respectively. Figure \textbf{18} indicates that the discrete red points always lie on the blue line at the smaller values of the horizon temperature and then gradually move away from it with increasing values of temperature in all cases. Whereas from Fig. \textbf{19}, it is noted that the discrete red points always lie above and below or closely to the blue line at the smaller or larger values of the horizon temperature. Hence, the change in the positions of the photon orbit and the Einstein radius obtained from the wave optics are observed with respect to different values of parameter $\alpha$ and $\omega$ in both figures. As expected, in all cases, the Einstein ring radius obtained by our wave optics fits well with that of geometric optics. Finally, we concluded that the holographic Einstein ring provides us with immense information about the geometric structure of space-time and helps us to analyze different gravity models comprehensively. Moreover, holographic images can be used as an effective tool to distinguish different types of BHs for fixed wave source and optical system. We hope these observations may provide a more intuitive understanding of the Einstein rings and their related consequences, for the tabletop experiments in the future.

\section*{Acknowledgements}

{This work is supported  by the National
Natural Science Foundation of China (Grants Nos. 11675140, 11705005,   12375043), Innovation and Development Joint  Foundation of Chongqing Natural Science  Foundation (Grant No. CSTB2022NSCQ-LZX0021) }, and Basic Research Project of Science and Technology Committee of Chongqing (Grant No. CSTB2023NSCQ-MSX0324).

\end{document}